\documentclass[12pt]{amsart}
\usepackage{amssymb,amsmath}
\usepackage{latexsym}
\usepackage{graphicx}
\usepackage{geometry}                
\usepackage{color}
\usepackage{hyperref}
\hypersetup{
    colorlinks=true, 
    linktoc=all,     
    linkcolor=hyptxt,  
}
\definecolor{hyptxt}{rgb}{0.5, 0, 0.5}
\definecolor{hyp}{rgb}{0.281,0.275,0.485}
\usepackage{bibtopic}
\newcommand{\mc}{\mathcal}
\newcommand{\Hil}{\mc H}
\newcommand{\C}{\mathbb C}

\newcommand{\N}{\mathbb N}

\newcommand{\R}{\mathbb R}

\def\lg{\langle }
\def\rg{\rangle }

\def\adg{a^{\dag}}

\def\deq{\stackrel{\mathrm{def}}{=}}

\def\lu{\mbox{\large 1}}

\begin{document}
\title[Quantizations with complex Hermite polynomials]{Quantizations from reproducing kernel spaces}

\author{S. Twareque Ali, Fabio Bagarello, and Jean Pierre Gazeau}
\address{Department of Mathematics and Statistics,
      Concordia University, Montr\'eal, Qu\'ebec, Canada H3G 1M8} \email{stali@mathstat.concordia.ca}
\address{Dieetcam, Facolt\`a di Ingegneria, Universit\`a di Palermo, I-90128 Palermo, Italy}\email{fabio.bagarello@unipa.it}
\address{Laboratoire APC, Universit\'e Paris 7-Denis Diderot, 10, rue A. Domon et L. Duquet, 75205 Paris Cedex13, France}\email{gazeau@apc.univ-paris7.fr}
\date{\today}
\begin{abstract}
The purpose of this work is to explore the existence and properties of reproducing kernel Hilbert subspaces of $L^2(\C, \,  d^2z/\pi)$ based on subsets of complex Hermite polynomials. The resulting coherent states (CS)  form a family depending  on a nonnegative parameter $s$.  We examine some  interesting issues, mainly related to CS quantization, like the  existence of  the usual harmonic oscillator spectrum despite the absence of canonical commutation rules. The question of mathematical and physical equivalences between the $s$-dependent quantizations is also considered.
\end{abstract}

\maketitle
\tableofcontents
\section{Introduction}
\label{intro}
It is well known that in  hamiltonian mechanics and with appropriate units the complex plane
\begin{equation}
\label{compplane}
\C = \left\lbrace z = \frac{q + ip}{\sqrt{2}} \right\rbrace
\end{equation}
represents the phase space for  the motion of a particle on the line. The symplectic measure on it  is Lebesgue, $d^2z/\pi$. A, say \textit{classical}, statistical reading of such a measure space rests upon the metric structure of the Hilbert space $L^2(\C, d^2z/\pi)$ of square integrable complex-valued classical observables $f(q,p)$ viewed as $(q,p)$-dependent \textit{signals}, i.e. images. The most straightforward way to obtain the quantum version of this phase space is to implement its so-called Berezin-Klauder-Toeplitz or ``anti-Wick'' quantization, judged as equivalent to canonical quantization on a physical level. This procedure rests upon the resolution of the unity produced by standard (i.e. Glauber-Sudarshan) coherent states  in the Fock-Bargman-Segal space $\mathcal{FBS}$,
\begin{equation}
\label{fockbarg}
\mathcal{FBS} = \left\lbrace \phi(z)= e^{-\vert z\vert^2/2}\,s(z)\in L^2(\C, d^2z/\pi)\, , \, s(z)\, \mbox{entire analytical}  \right\rbrace
\end{equation}
The space $\mathcal{FBS}$ is a reproducing kernel Hilbert subspace of $L^2(\C, d^2z/\pi)$ with kernel $\mathrm{K}(z,\overline{z'}) = e^{ z \overline{z'}}$, $ \phi(z) = \int_{\C}\frac{d^2 z}{\pi} \mathrm{K}(z,\overline{z'})\, \phi(z')$, and the function $z\mapsto \zeta_{\overline{z'}}(z)= e^{-\vert z\vert^2/2}\,e^{z  \overline{z'}}$  is  a coherent state in its
Fock-Bargman representation. The  simplicity of quantum mechanics, specially  its beautiful Weyl-Heisenberg symmetry encoded by the CCR $[Q,P]= i \lu$, where $Q = (a + a^{\dag})/\sqrt{2}$, $P= -i(a - a^{\dag})/\sqrt{2}$, $a = \partial/\partial z + \bar z/2$, and $a^{\dag}= -\partial/\partial \bar z +  z/2$, derives from this underlying analytic structure.

Now, if we explore more thoroughly the ``large'' classical arena $L^2(\C, d^2z/\pi)$, we find an unsuspected richness which goes far beyond this simplest Fock-Bargman-Segal subspace. The  richness rests upon the existence of a specific orthonormal basis built from complex Hermite polynomials, introduced recently by Ghanmi  in \cite{ghanmi08} and lately explored within a quantum mechanical context in \cite{alibaga10} and \cite{cotgazgor10}. This ``large'' basis can be partitioned in an infinity of sectors leading to an (almost direct) sum decomposition of $L^2(\C, d^2z/\pi)$ into reproducing kernel Hilbert subspaces, denoted in this paper by $\mathcal{K}^{\epsilon}_s$, $s\in \{0,1,2,\dotsc\} \equiv \N$, and $\epsilon = L$ (for ``left'') or $R$ (for ``right''). The most remarkable feature, already exploited in \cite{cotgazgor10}, is the possibility offered by such a family of subspaces  of studying a $s$-labeled set of quantizations.  These latter quantizations are similar to that inherent to Fock-Bargman-Segal quantization, possibly even equivalent, or non-equivalent to it, depending on the definition we agree to give to the adjective ``equivalent'', on a physical, observational level, or on a purely mathematical level. It is also intriguing to observe the apparently simple (but heavy in consequences!) modification of the CCR: $[a_s,a_s^{\dag}] = \lu + s P_0$, where $P_0$ is the orthogonal projector on the lowest state in the considered Hilbert space  which naturally arises here.

The aim of this paper is to go forward and deeper in the investigation of these Hilbert space and quantization aspects.   Section \ref{comherm} is a short review of the definition and properties of complex Hermite polynomials. The resulting decomposition of the space $L^2(\C, d^2z/\pi)$ into subspaces and the ladder operators allowing one to pass from one subspace to a contiguous one
are described in Section \ref{decladder}. The appearance in this ladder algebra formalism of non-linear pseudo-bosons \cite{bagnlpb} is explained in Section \ref{pseudoboson}. Then families of coherent states built from complex Hermite polynomials are introduced in Section \ref{CScomherm}.
We then proceed in Section \ref{CScomhermquant} with the complex Hermite CS quantization of the complex plane and functions on it. Some crucial functional properties of the  resulting position and momentum operators are examined in  Section \ref{chihara}, in particular the study of the $s$-labeled families of associated orthogonal polynomials which extend the ordinary Hermite polynomials appearing in the $s=0$ case. After restoring physical dimensions, namely a mass $m$, the Planck constant $\hbar$ and a fundamental length (e.g. the Compton length) we consider in Section \ref{canvcs} the question of physical equivalence between the elements of this $s$-labeled family of quantizations with regard to the spectrum of the quantum harmonic oscillator.
We end the paper in Section \ref{seconc} with a few remarks on the interest of such explorations, and technical details are given in appendices.

\section{Complex Hermite polynomials: definition and properties}
\label{comherm}
Let $r$ and $s$ be nonnegative integers, i.e. $r\,,\,s\in \N$. Complex Hermite polynomials are defined as \cite{ghanmi08} (see also \cite{ito}):
\begin{equation}\label{eqCH1}
h^{r,s}(z,\bar{z}) = (-1)^{r+s}\, e^{\vert z \vert^2}\, \frac{\partial^r}{\partial z^r}\, \frac{\partial^s}{\partial \bar z^s}\, e^{-\vert z \vert^2} = \sum_{k=0}^{\inf(r,s)} \frac{(-1)^k}{k!}\,\frac{r!s!}{(r-k)!(s-k)!}\, z^{s-k}\, \bar z^{r-k}\, .
\end{equation}
An immediate consequence of this definition is their symmetry with respect to index permutation:
\begin{equation}
\label{hermperm}
h^{r,s}(z,\bar{z})  = \overline{h^{s,r}(z,\bar{z}) }\, .
\end{equation}
They form a complete orthogonal system in the Hilbert space $L^2\left(\C\,,\, e^{-\nu\vert z \vert^2} \,d^2z\right)$ with $\nu > 0$.

We now suppose that $r \geq s$. Then the corresponding  polynomials can be written  in terms of confluent hypergeometric functions or in terms of associate Laguerre polynomials:
\begin{eqnarray}\nonumber
h^{s+n, s}(z,\bar{z}) &=& s! (s+n)!\, \bar{z}^{n}\, \sum_{k = 0}^{s} \frac{(-1)^{s-k}}{(s-k)!}\,\frac{\vert z\vert^{2k}}{k!\,(k+n)!}, \\[0.5\baselineskip] \label{eqCH2}
\nonumber&=& \frac{(-1)^{s} \,(s+n)!}{n!} \, \bar{z}^{n} \,_{1}F_{1}(-s; n + 1; \vert z\vert^{2}) \\
&=&  (-1)^{s} \,s!\, \bar{z}^{n} \, L_s^{(n)}(\vert z\vert^{2}) \, ,
\end{eqnarray}
where $r-s=n\in\mathbb{N}$. In particular, for $s=0$ and $1$, the expression  (\ref{eqCH2}) reduces, respectively,  to $\bar{z}^{n}$ and $\bar{z}^{n}(\vert z\vert^{2} - n-1)$, and for $n=0$ it reads as:
\begin{equation}
\label{hermn0}
h^{s, s}(z,\bar{z})= (-1)^{s} \,s! \, L_s^{(0)}(\vert z\vert^{2})\equiv  (-1)^{s} \,s! \, L_s(\vert z\vert^{2})\, .
\end{equation}

 For a fixed $s$ we have an infinite  family of pairwise orthogonal complex polynomials  of degree $n+2s$ in variables $z$ and $\bar z$. Precisely, by using the relation (2.20.1.19) in \cite{Prud1}, we obtain:
\begin{equation}\label{eqCH3}
\frac{1}{\pi}\,\int_{\C} d^{2}z\, e^{-\vert z\vert^{2}}\, h^{s+n, s}(z, \bar z) \overline{h^{s+n', s}(z, \bar z)} = \left\{
\begin{array}{lll}
s!\, (s+n)! & {\rm if} & n= n'\\[2mm]
0 & {\rm if} & n\neq n'\,.
\end{array} \right.
\end{equation}
For any pair $s$, $s'$,  the orthogonality  of $h^{s+n, s}$ and  $h^{s'+n', s'}$ for $n\neq n'$ results from the angular part integration in the complex plane, whereas at $n=n'$, the orthogonality for $s\neq s'$ results from the orthogonality of the generalized Laguerre polynomials:
\begin{equation}
\label{lagorthog}
\int_0^{\infty}e^{-t}\, t^{\alpha}\, L_s^{(\alpha)}(t)\, L_{s'}^{(\alpha)}(t)\, dt = \delta_{s s'}\, \Gamma(1+\alpha)\, \binom{s+\alpha}{s}\, .
\end{equation}
{\color{red} }
The functions $h^{s+n,s}$ are related through the ladder operators
\begin{eqnarray}\label{hermlowrais} \left\{
\begin{array}{l}
\left( -\dfrac{\partial }{\partial z}+\bar z\right) h^{s+n,s}=h^{s+n+1,s} \\[5mm]
\dfrac{\partial }{\partial \bar z}h^{s+n+1,s}=(s\!+\!n\!+\!1)\, h^{s+n,s}
\end{array}
\qquad \right\{
\begin{array}{l}
\left( -\dfrac{\partial }{\partial \bar z}+z\right) h^{s+n,s}=h^{s+n,s+1}\, ,  \\[5mm]
\dfrac{\partial }{\partial z}h^{s+n,s+1}=(s\!+\!1)\, h^{s+n,s}\, .
\end{array}
\end{eqnarray}

\section{Orthonormal basis of $L^2( \C, d^2z/\pi)$ and displacement operator}
\label{obdisp}
\subsection{Orthonormal basis}
Let us fix $s$ and introduce the Hilbert subspace $\mathcal{K}^L_s$ (resp.  $\mathcal{K}^R_s$) in $L^2( \C, d^2z/\pi)$, subscript $L$  (resp. $R$)  standing for ``left''  (resp. ``right''),  as the closure of the linear span of the set of orthonormal functions $\phi^L_{n;s}$ (resp. $\phi^R_{n;s}$) defined as
\begin{equation} \label{phisn}
\begin{array}{ll}
\nonumber \phi^L_{n;s}(z,\bar z) = \overline{\phi^R_{n;s}}(z,\bar z)&\deq \dfrac{1}{\sqrt{s! (s+n)!}}e^{-\vert z\vert^{2}/2}\, h^{s+n, s}(z, \bar z)\, , \\
& = (-1)^{s}  \sqrt{\dfrac{s!}{(s+n)!}}\,e^{-\vert z\vert^{2}/2}\,\bar{z}^{n} \, L_s^{(n)}(\vert z\vert^{2})\, .
\end{array}
\end{equation}
In particular we note that $\phi^L_{0;s}(z,\bar z)  = \phi^R_{0;s}(z,\bar z) \equiv \phi_{0;s}(z,\bar z) = (-1)^s e^{-\vert z\vert^{2}/2}\, L_s(\vert z\vert^{2})$ for all $s\in \N$.
\subsection{Its Weyl-Heisenberg origin}
The orthonormality of the family of the  above functions (and the related orthogonality  the complex Hermite polynomials) is easily understood from they are, up to a phase factor, matrix elements of the unitary Weyl-Heisenberg displacement operator $D(z)$ with respect to the Fock number  basis $|n\rg$, $n\in \N$. Recall that this operator is defined as
\begin{equation}
\label{displac}
D(z) = e^{z\adg -\bar z a}\, ,\quad D(-z) = (D(z))^{-1} = D\dag(z)\, ,
\end{equation}
where $a|n\rg = \sqrt{n}|n-1\rg$, $\adg|n\rg = \sqrt{n+1}|n+1\rg$, $a|0\rg = 0$, $[a,\adg]= \mbox{\large 1}$.
One can find in \cite{cahillglauber69} an exhaustive list of properties of $D(z)$. In particular, its matrix elements $D_{ms}(z)$ in the number basis are simply related, for $n = m-s\geq 0$, to the functions (\ref{phisn}) by:
\begin{equation}
\label{relDPhi}
\lg n+s|D(z)|s\rg = D_{n+s s}(z) = (-1)^s \phi^R_{n;s}(z,\bar z)= (-1)^s \overline{\phi^L_{n;s}(z,\bar z)}\,.
\end{equation}
Orthonormality properties straightforwardly derive from unitarity:
\begin{equation}
\label{unitortho}
\int_{\C}\frac{d^2 z}{\pi}\, D_{mn}(z)\, \overline{D_{m'n'}(z)}= \delta_{mm'}\delta_{nn'}\,.
\end{equation}
Moreover, one derives from unitarity the infinite sums:
\begin{equation}
\label{sumser}
\sum_{n=0}^{\infty}D_{mn}(z)\overline{D_{m'n}(z)}= \delta_{mm'}= \sum_{n=0}^{\infty}D_{nm'}(z)\overline{D_{nm}(z)}\, ,
\end{equation}
and particularly
\begin{equation}
\label{sumser2}
\sum_{n=0}^{\infty} \vert D_{mn}(z)\vert^2= 1= \sum_{n=0}^{\infty} \vert D_{nm}(z)\vert^2 \, , \quad m \in \N\, .
\end{equation}
The following important  inequality is then derived from (\ref{relDPhi}) and (\ref{sumser2}):
\begin{equation}
\label{ineqphi}
\sum_{n=0}^{\infty} \vert \phi^{R\,\mathrm{or}\,L}_{n;s}(z,\bar z) \vert^2 = \sum_{m\geq s}^{\infty} \vert D_{ms}(z)\vert^2< 1 \, , \quad s=1,2,\dotsc  \, .
\end{equation}

\section{Ladder operators and decomposition of $L^2( \C, d^2z/\pi)$}
\label{decladder}

Let us introduce the following  four operators in $L^2( \C, d^2z/\pi)$
\begin{align}
\label{ALRK}
A^L &=    \frac{\partial }{\partial \bar z}+\frac{z}{2}\, , \quad {A^L}^{\dag}=   -\frac{\partial }{\partial z}+\frac{\bar z}{2} \, , \\
    A^R &=    \frac{\partial }{\partial  z}+\frac{\bar z}{2}\, , \quad {A^R}^{\dag}=  -\frac{\partial }{\partial \bar z}+\frac{ z}{2} \, .
  \end{align}
  Together with the identity, they are generators of two independent (mutually commuting) Weyl-Heisenberg algebras:
  \begin{equation}
\label{WHLR}
[A^L, {A^L}^{\dag}] = \mbox{\large 1}\, , \quad [A^R, {A^R}^{\dag}] = \mbox{\large 1}\, .
\end{equation}
Due to the equations in  (\ref{hermlowrais}), the functions $\phi^L _{n;s}$ and $\phi^R _{n;s}$ are related through these ladder operators
\begin{align}
\label{ladlowupL}
A^L\phi^L _{n+1; s}&=\sqrt{s+n+1}\, \phi^L _{n; s}\, , \quad {A^L}^{\dag}\phi ^L_{n; s}=\sqrt{s+n+1}\, \phi^L _{n+1; s}\, , \\
\label{ladlowupR} A^R\phi^R _{n+1; s}&=\sqrt{s+n+1}\, \phi^R _{n; s}\, , \quad {A^R}^{\dag}\phi ^R_{n; s}=\sqrt{s+n+1}\, \phi^R _{n+1; s}\, .
\end{align}
We thus obtain  a countably infinite family of Hilbert subspaces $\mathcal{K}^L_s$ and $\mathcal{K}^R_s$, $s \in \N$. At a given $s$,  $\mathcal{K}^L_s$ and $\mathcal{K}^R_s$ share their ``lowest'' one-dimensional subspace:
\begin{equation}
\label{lowdimint}
G_s \deq \mathcal{K}^L_s\cap\mathcal{K}^R_s = \{ \lambda \phi _{0; s}\,, \, \lambda\in \C  \}\, .
\end{equation}
The ``canonical'' anti-Fock-Bargmann (for $L$)  and Fock-Bargmann (for $R$) subspaces correspond to $s=0$. They share  the ``absolute''  ground state $\phi^L _{0; 0}=  \phi^R _{0; 0} \equiv \phi _{0}  $.  Note that, at the exception of the latter  cases for which $A^L\phi _{0}= 0 = A^R\phi _{0}$,  the  lowest states $\phi _{0; s}$, $s>0$,  are not cancelled by $A^L$ and $A^R$. We have instead the following relations:
\begin{equation}
\label{lowgrstats}
A^L\phi _{0; s} = \sqrt{s}\,\phi^R_{1; s-1} \,, \quad A^R\phi _{0; s}=  \sqrt{s}\,\phi^L_{1; s-1}\, , \ s = 1, 2, \dotsc\, .
\end{equation}

More generally,  the action of right (resp. left)  ladder operators on left (resp. right) subspaces read as:
\begin{align}
\label{ladLR}
  A^R \phi^L _{n; s} &= \sqrt{s} \, \phi^L _{n+1; s-1} \,, \quad   A^L \phi^R _{n; s} = \sqrt{s} \, \phi^R _{n+1; s-1}\, , \\
      {A^R}^{\dag} \phi^L _{n; s} &= \sqrt{s+1} \, \phi^L _{n-1; s+1} \,, \quad   {A^L}^{\dag} \phi^R _{n; s} = \sqrt{s+1} \, \phi^R _{n-1; s+1} \, .
\end{align}
In particular, $A^R$ (resp. $A^L$) annihilates the anti-Fock-Bargmann (resp. Fock-Bargmann) subspace:
\begin{equation}
\label{annihFB}
A^R \mathcal{K}^L_0 =0\, , \quad  A^L\mathcal{K}^R_0 = 0\, .
\end{equation}
The above results are in agreement with the fact that, calling for instance $N^L={A^L}^{\dag}{A^L}$, $N^L\phi^L _{n; s}=(n+s)\phi^L _{n; s}$ and $[N^L,A^R]=[N^L,{A^R}^{\dag}]=0$.

Let us now define the subspaces $ \mathcal{K}^L_{\ast s}$ and $ \mathcal{K}^R_{\ast s}$ by:
\begin{equation}
\label{ksup}
\mathcal{K}^L_{\ast s} = \mathcal{K}^L_s\setminus G_s\, , \quad \mathcal{K}^R_{\ast s} = \mathcal{K}^R_s\setminus G_s\, .
\end{equation}
We then have the following direct sum (orthogonal) decomposition of $L^2( \C, d^2z/\pi)$:
\begin{equation}
\label{orthdec}
L^2( \C, d^2z/\pi) = \bigoplus_{s=0}^{\infty} \left(  \mathcal{K}^L_{\ast s}\oplus  G_s\oplus \mathcal{K}^R_{\ast s}\right) \equiv \mathcal{L}^L \oplus \mathcal{L}_0 \oplus \mathcal{L}^R\, ,
\end{equation}
in which $\mathcal{L}^L =  \oplus_{s=0}^{\infty}  \mathcal{K}^L_{\ast s}$, $\mathcal{L}^R =  \oplus_{s=0}^{\infty}  \mathcal{K}^R_{\ast s}$, and $\mathcal{L}_0 =  \oplus_{s=0}^{\infty} G_s.$

The organization of this decomposition can be considered to be natural with regard to complex conjugation or mirror symmetry $J:L\mapsto R$ with respect to the ``hyperplane'' $\mathcal{L}_0$:
\begin{equation}
\label{mirsymJ}
J\, \phi^L _{n; s} = \phi^R _{n; s}= \overline{ \phi^L _{n; s}}\, , \ \, J\, \phi_{0; s} = \phi _{0; s}\, , \quad J^2 = \mbox{\large 1}\, .
\end{equation}
Ladder operators for the subspace $\mathcal{L}_0$ of absolute ($s=0$) and relative ($s>0$) ground states are quadratic, mixing left and right, and are found from the following equations:
\begin{equation}
\label{grlad}
A^L\,A^R\phi _{0; s}=s\, \phi _{0; s-1}\, , \quad  {A^R}^{\dag}  {A^L}^{\dag}\phi _{0; s}=(s+1)\, \phi _{0; s+1}\, .
\end{equation}

\section{First consequences of this setting}
\label{pseudoboson}
The operators introduced so far allow us to construct, first of all, examples on non-linear pseudo-bosons (NLPB), \cite{bagnlpb}. For the sake of completeness, we recall here that the NLPB are defined as a triple $(a,b,\{\epsilon_n\})$ where $a$ and $b$ are two operators on the Hilbert space $\Hil$, and $\{\epsilon_n\}$ is a sequence of positive numbers such that $0=\epsilon_0<\epsilon_1<\cdots<\epsilon_n<\cdots$, satisfying the following conditions:
{\bf (p1)} a non zero vector $\varPhi_0$ exists in $\Hil$ such that $a\,\varPhi_0=0$ and $\varPhi_0\in D^\infty(b)$;
{\bf (p2)} a non zero vector $\varPsi_0$ exists in $\Hil$ such that $b^\dagger\,\varPsi_0=0$ and $\varPsi_0\in D^\infty(a^\dagger)$;
{\bf  (p3)} calling
$
\varPhi_n:=\frac{1}{\sqrt{\epsilon_n!}}\,b^n\,\varPhi_0$, and $\varPsi_n:=\frac{1}{\sqrt{\epsilon_n!}}\,{a^\dagger}^n\,\varPsi_0,$
we have, for all $n\geq0$,
$
a\,\varPhi_n=\sqrt{\epsilon_n}\,\varPhi_{n-1}$, and $b^\dagger\varPsi_n=\sqrt{\epsilon_n}\,\varPsi_{n-1}
$;
{\bf  (p4)} the sets ${\mc F}_\varPhi=\{\varPhi_n,\,n\geq0\}$ and ${\mc F}_\varPsi=\{\varPsi_n,\,n\geq0\}$ are bases of $\Hil$.

\vspace{1mm}

Here $D^\infty(X)$ is the domain of all the powers of the operator $X$. Let us now introduce $a=A^LA^R\,N^L$, $b={A^R}^\dagger{A^L}^\dagger$ and $\epsilon_n=n^3$. Then, taking $\varPhi_0=\varPsi_0=\phi _{0; 0}$, it is easy to check that the conditions above are all satisfied. In particular, for instance, we find $\varPhi_n=\frac{1}{\sqrt{n!}}\,\phi _{0; n}$ and $\varPsi_n=\sqrt{n!}\,\phi _{0; n}$, which incidentally shows that ${\mc F}_\varPsi$ is not a Riesz basis. Hence our NLPB are not regular, \cite{bagnlpb}. We recall that $\varPhi_n$ and $\varPsi_n$ are eigenstates of the (in general) non-self adjoint operators $M=ba$ and $M^\dagger=a^\dagger b^\dagger$, both with eigenvalue $\epsilon_n$. As a matter of fact, for our choices of $a$ and $b$, we get $M=M^\dagger=N^R{N^L}^2$. This choice also implies that the intertwining operators $S_\varPhi=\sum_n|\varPhi_n\rg\lg \varPhi_n|$, $S_\varPsi=\sum_n|\varPsi_n\rg\lg\varPsi_n|$, which in general satisfy $MS_\varPhi=S_\varPhi M^\dagger$,  commute with $M$: $[S_\varPhi,M]=[S_\varPsi,M]=0$.

\vspace{3mm}

Another interesting aspect of the settings discussed in the first part of this section is that the two operators $N^R$ and $N^L$ can be viewed essentially as the two {\em dual} hamiltonian operators arising from the procedure introduced in \cite{bagint}, which we here briefly recall: let $h_1$ be a self-adjoint hamiltonian on  $\Hil$, $h_1=h_1^\dagger$, whose normalized eigenvectors, $\hat\varphi_n^{(1)}$, satisfy the following equation: $h_1\hat\varphi_n^{(1)}=\epsilon_n\hat\varphi_n^{(1)}$, $n\in\Bbb{N}$. Suppose that there exists an operator $x_1$ on $\Hil$ such that $
[x_1x_1^\dagger,h_1]=0,
$
and  $N:=x_1^\dagger\,x_1$ is invertible.  Then, calling
$h_2:=N^{-1}\left(x_1^\dagger\,h_1\,x_1\right)$  and $\varphi_n^{(2)}=x_1^\dagger\hat\varphi_n^{(1)},
$
we find that
$h_2=h_2^\dagger$,
$ x_1^\dagger\left(x_1\,h_2-h_1\,x_1\right)=0$ and, if
$\varphi_n^{(2)}\neq 0$, then $h_2\varphi_n^{(2)}=\epsilon_n\varphi_n^{(2)}.$

Let now take $h_1=N^R$ and $x_1=J{A^L}^\dagger$. Then our requirements are satisfied and we find that $h_2=A^L{A^L}^\dagger=N^L+\mbox{\large 1}$. Moreover, since $\varphi_n^{(1)}=\phi^R_{n; s}$, then $\varphi_n^{(2)}=A^LJ\phi^R_{n; s}=\sqrt{n+s}\,\phi^L_{n-1; s}$, for $n\geq1$. It is easy to check that $h_2\varphi_n^{(2)}=(n+s)\varphi_n^{(2)}$, that $h_2=h_2^\dagger$, and that $x_1^\dagger\left(x_1\,h_2-h_1\,x_1\right)=A^L{A^L}^\dagger-A^LN^L{A^L}^\dagger=0$. We can also check easily that $[h_2,x_1^\dagger x_1]=0$.

\vspace{3mm}

It is not hard to imagine that several other examples of dual hamiltonians  $(h_1,h_2)$, and of NLPB, can be constructed out of the framework described in the previous section. {These results are of a certain interest particularly in connection with} two quite recent {\em hot} generalizations of quantum mechanics (q.m.), i.e. to supersymmetric q.m. and to crypto (or pseudo)-hermitian q.m., \cite{mosta,bender,zno}, and all its variations.

\section{Complex Hermite coherent states}
\label{CScomherm}

\subsection{Coherent state construction: a guideline}
\label{seccsquant}

We give here a short account of a general method for building (overcomplete) families of unit-norm states resolving the identity in some Hilbert space (see \cite{gazeaubook09} for more details).

Let $\Sigma$ be a set of parameters equipped with a measure $\mu$ and its associated Hilbert space $L^2(\Sigma, \mu)$ of complex-valued square integrable functions with respect to $\mu$. Let us choose
in $L^2(\Sigma, \mu)$ a finite or countable  orthonormal set $\mathcal{O}=\{\phi_n\, , \, n = 0, 1, \dots \}$:
\begin{equation}\label{eqI1}
\lg \phi_m | \phi_n \rg = \int_{\Sigma}\overline{\phi_m(\alpha)}\, \phi_n(\alpha)\, \mu(d\alpha) = \delta_{mn}\, .
\end{equation}
In case of infinite countability, this set must obey the (crucial) positiveness and finiteness conditions:
\begin{equation}\label{eqI2}
0<\sum_{n} \vert \phi_n (\alpha)\vert^2 \deq \mathcal{N}(\alpha) < \infty \,  \quad \mathrm{a.e.}
\end{equation}
Let $\mathcal{H}$ be a separable complex Hilbert space with orthonormal basis $\{|e_n\rg\, , \, n = 0, 1, \dots \}$ in one-to-one correspondence with the elements of  $\mathcal{O}$. From Conditions (\ref{eqI1}) and (\ref{eqI2}) there results that the  family of normalized ``coherent'' states $\mathcal{F}_{\mathcal{H}}= \{|\alpha\rg\, , \, \alpha \in \Sigma \}$ in $\mathcal{H}$, which are defined by
\begin{equation}\label{eqI3}
|\alpha\rg = \frac{1}{\sqrt{\mathcal{N}(\alpha)}}\sum_n \overline{\phi_n(\alpha)}\, |e_n\rg\, ,
\end{equation}
resolves the identity in $\mathcal{H}$:
\begin{equation}\label{eqI4}
\int_\Sigma \mu(d\alpha) \,\mathcal{N}(\alpha) \, |\alpha\rg \lg \alpha | = \mbox{\large 1}_{{\mathcal H}}\, .
\end{equation}
Considering the isometric map ${\mathcal H} \ni \psi \mapsto  \sqrt{\mathcal{N}(\alpha)}\lg \alpha|\psi\rg \deq \phi(\alpha) \in L^2(\Sigma, \mu(d\alpha))$, the closure of the image of $\mathcal{H}$ is a reproducing Hilbert space, subspace of $L^2(\Sigma, \mu(d\alpha))$, with reproducing kernel $\mathrm{K}(\alpha,\alpha') = \sqrt{\mathcal{N}(\alpha)}\lg \alpha|\alpha'\rg
\sqrt{\mathcal{N}(\alpha')}= \sum_{n} \phi_n (\alpha)\, \overline{\phi_n (\alpha')} =\overline{\mathrm{K}(\alpha',\alpha) }$:
\begin{equation}
\label{repkereq}
\phi(\alpha) = \int_{\Sigma} \mu(d\alpha')\, \mathrm{K}(\alpha,\alpha') \phi(\alpha')\, .
\end{equation}

\subsection{Using complex Hermite polynomials}
\label{CScomhermA}

To each fixed $s$ there corresponds the two Hilbert subspaces, $\mathcal{K}^{\epsilon}_s$, $\epsilon = L$ or $R$, together with their respective orthonormal bases $\{\phi^{\epsilon} _{n; s}\, , \, n \in \N\}$. Following the guideline indicated in Section \ref{seccsquant}, we consider the set $X = \C$ equipped with the Lebesgue (or ``uniform'') measure $\mu(dx)= d^2z/\pi$, and, for a fixed $s$, we choose the orthonormal set $\mathcal{O} = \{\phi^{\epsilon} _{n; s}\, , \, n \in \N\}$ in $L^2( \C, d^2z/\pi)$.   We  then construct the coherent states based on complex Hermite polynomials  as the  infinite linear combination of orthonormal elements $|e^{\epsilon}_n;s\rangle$  of some separable Hilbert space $\mathcal{H}^{\epsilon}_{s}$:
\begin{equation}\label{eqCS0}
|z;s, {\epsilon}\rangle = \frac{1}{\sqrt{e^{-\vert z \vert^2}\,\mathcal{N}_{s}(\vert z\vert^{2})}}\,\sum_{n=0}^{\infty}\,\overline{\phi^{\epsilon}_{n;s}(z)}|e^{\epsilon}_n;s\rangle   \, .
\end{equation}
Note the change of notation compared to Eq. (\ref{eqI2}), necessary in order to eliminate a Gaussian factor.
More precisely,
\begin{equation}\label{eqCS1}
|z; s;L\rangle \,=\, \frac{(-1)^{s}}{\sqrt{\mathcal{N}_{s}(\vert z\vert^{2})}} \, \sum_{n=0}^{\infty}\,
{\left(
\begin{array}{c}
s\!+\!n\\
s
\end{array}
\right)}^{-1/2}\,
\frac{z^{n}}{\sqrt{n!}} \,L_s^{(n)}(\vert z\vert^{2}) \, |e^L_n;s\rangle\, ,
\end{equation}
\begin{equation}\label{eqCS1R}
|z; s;R\rangle \,=\, \frac{(-1)^{s}}{\sqrt{\mathcal{N}_{s}(\vert z\vert^{2})}} \, \sum_{n=0}^{\infty}\,
{\left(
\begin{array}{c}
s\!+\!n\\
s
\end{array}
\right)}^{-1/2}\,
\frac{\bar z^{n}}{\sqrt{n!}} \,L_s^{(n)}(\vert z\vert^{2}) \, |e^R_n;s\rangle\, .
\end{equation}

One can choose all spaces $\mathcal{H}^{\epsilon}_{s}$  as identical, e.g. the Fock space spanned by number states $|n\rg$, or the Hilbert space $L^2(\R, dx)$, in which case there is no need to specify the parameter $s$. Another choice could be  $\mathcal{H}^{\epsilon}_{s} = \mathcal{K}^{\epsilon}_{s}$ and then one identifies  the states $|e^{\epsilon}_n;s\rg$ with the functions $\phi^{\epsilon}_{n;s}$.

The normalization factor is defined as
\begin{align}\nonumber
\mathcal{N}_{s} (\vert z \vert^2)\,&=\, \sum_{n = 0}^{\infty}\frac{\vert h^{s+n,s}(z, \bar z)\vert^{2}}{s!(s+n)!}= \, \sum_{n = 0}^{\infty} \frac{s!}{ (s+n)!} \, \vert z\vert^{2n}\,\left(L_s^{(n)}(\vert z\vert^{2}) \right)^2\\
\nonumber  & = \sum_{m,m'=0}^s (-1)^{m+m'}\,\frac{(s!)^2}{(s-m)! (s-m')! (m!)^2 (m'!)^2}\, \vert z \vert^{2(m+m')}\times \\
\label{eqCS2} &\hskip 3.5cm \times {}_2F_2(1,s+1;m+1,m'+1;\vert z\vert^2)\,  .\end{align}

We easily check that for $s = 0$ (resp. $s=1$) the series (\ref{eqCS2}) reduces to  $e^{\vert z\vert^2}$ (resp. $e^{\vert z\vert^{2}}-\vert z\vert^{2}$). { For higher $s$ we already know from the upper bound (\ref{ineqphi}) that:
\begin{equation}
\label{ineqN}
\mathcal{N}_{s} (\vert z\vert^2) = e^t(1-\sum_{m=0}^{s-1} \vert D_{ms}(z)\vert^2 < e^t\, , \quad t=\vert z\vert^2\, ,
\end{equation}
which insures that the requirement (\ref{eqI2})  is satisfied. Moreover, from the expression of the matrix elements $D_{mn}$ we
obtain the expression:
\begin{equation}
\label{behN(t)}
\mathcal{N}_{s} (t) = e^t - \sum_{m=0}^{s-1}\frac{m!}{s!} t^{s-m}\left(L^{(s-m)}_m(t)\right)^2= e^t - Q_{2s-1}(t)\, ,
\end{equation}
where  $Q_{2s-1}$ is polynomial of degree $2s-1$ such that $Q_n(0) = 0$. }

For $s=0$ states (\ref{eqCS0}) are the standard coherent states or their conjugate version, but for the remaining values of $s$ we are in presence of  some deformation of the standard $|z; 0\rangle \equiv |z\rg$. Therefore we have with Eqs.\,(\ref{eqCS1},\ref{eqCS1R})  an infinite family of coherent states families, which is labeled by $s\in \N$ and by $\epsilon$. By construction these states are unit vectors and they resolve the unity in their respective spaces:
\begin{equation}
\label{resunity1}
\int_{\C} \frac{d^2z}{\pi} \,e^{-\vert z \vert^2}\,  \mathcal{N}_s(\vert z\vert^2) \, | z;s, {\epsilon} \rg \lg z;s, {\epsilon} | = \mbox{\large 1}_{{\mathcal H}_s^{\epsilon}}\, .
\end{equation}

As noted in Subsection \ref{seccsquant} the resolution of the identity goes with the existence of  a reproducing kernel with corresponding reproducing Hilbert spaces $\mathcal{K}^{\epsilon}_{s}$:
\begin{align}
\label{repker}
\nonumber \mathrm{K}^{L}_s(z,\overline{z'}) &= e^{-\vert z \vert^2/2}\,  \sqrt{\mathcal{N}_s(\vert z\vert^2)} \, \lg z;s, L | z';s, L \rg   e^{-\vert z' \vert^2/2}\,  \sqrt{\mathcal{N}_s(\vert z'\vert^2)}\\
& =  \sum_{n = 0}^{\infty} \frac{s!}{ (s+n)!} \, (\bar z z')^{n}\,L_s^{(n)}(\vert z\vert^{2}) \,L_s^{(n)}(\vert z'\vert^{2}) \\
&= \mathrm{K}^{R}_s(z',\overline{z}) \, .
\end{align}
For $s=0$ we recover the simple exponential kernel $e^{\bar z z'}$. For $s=1$, we get the expression:
\begin{equation}
\label{repkers1}
\mathrm{K}^{L}_1(z,\overline{z'}) = e^{\bar z z'} (1- \vert z-z' \vert^2) - z\overline{z'}\, .
\end{equation}

\subsection{Bayesian probabilistic content}
\label{bayes}
There is, at the basis of the construction of coherent states outlined in {Subsection} \ref{seccsquant}, a deep Bayesian content \cite{algahel08}, which could possibly be based on experimental evidences, that is an interplay between the set of probability distributions $\alpha \mapsto \vert \phi_n(\alpha)\vert^2$ (from $\int_X \mu(d\alpha)\, \vert \phi_n(\alpha)\vert^2=1$), labelled by $n$,  on the classical measure space $(X,\mu)$, and the discrete set of probability distributions $n \mapsto \vert \phi_n(\alpha)\vert^2/\mathcal{N}(\alpha)$ (from $\mathcal{N}(\alpha)= \sum_n \vert\phi_n(\alpha)\vert^2$). Let us make explicit in the present context these two sets of probability. Since there is no angular dependence in the present situation, the only parameters or variable to be considered are $n$, $s$, and  $\vert z\vert^2 = t \in \R_+$:
\begin{enumerate}
  \item[(i)] the continuous distribution
  \begin{equation}
\label{baycont}
  t \mapsto  \frac{s!}{(s+n)!}\,e^{-t}\, t^n\, \left(L_s^{(n)}(t)\right)^2\, ,
\end{equation}
(for fixed $n$ and $s$)  with respect the measure $dt$, which generalizes the gamma distribution,
  \item[(ii)] the discrete distribution
   \begin{equation}
\label{baycont2}
  n \mapsto  \frac{s!}{(s+n)!}\, \frac{ t^n\, \left(L_s^{(n)}(t)\right)^2}{\mathcal{N}_s(t)}\, ,
\end{equation}
(for fixed $t$ and $s$) which generalizes the Poisson distribution.
\end{enumerate}

\section{Quantization with complex Hermite coherent states}
\label{CScomhermquant}
\subsection{Coherent state quantization: a short review}
The resolution of the identity (\ref{eqI4}) allows  to implement a \emph{coherent state or frame quantization} \cite{klauder63a,klauder63b,berezin75,klauder95,alienglis05,alietal09,gazeaubook09} of the set of parameters $\Sigma$ by associating to a function $\Sigma \ni \alpha \mapsto f(\alpha)$ that satisfies appropriate conditions the following operator in $\mathcal{H}$:
\begin{equation}\label{eqI5}
f(\alpha) \mapsto A_f  \deq \int_\Sigma\mu(d\alpha) \,\mathcal{N}(\alpha) \, f(\alpha)\, |\alpha\rg \lg \alpha |\, .
\end{equation}
Operator $A_f$ is symmetric if $f(\alpha)$ is real-valued, and is bounded (resp. semi-bounded) if $f(\alpha)$ is bounded  (resp. semi-bounded).  In particular, the Friedrich extension allows to define  $A_f$ as a self-adjoint operator  if $f(\alpha)$ is a semi-bounded real-valued function.  Note that the original $f(\alpha)$  is a ``upper symbol'', usually non-unique,  for the operator $A_f$. It will be called a
\emph{classical} observable with respect to the family $\mathcal{F}_{\mathcal{H}}$ if the so-called
``lower symbol" $\check{A}_f (\alpha)\deq\lg \alpha | A_f | \alpha \rg$ of $A_f$ has mild functional properties which can be made precise with additional topological properties imposed on the original set $\Sigma$.

\subsection{Hermite CS quantization}
We now proceed with the  quantization through the complex Hermite coherent states along the linear map (\ref{eqI5}):
\begin{align}
\label{hermquant1}
\nonumber f(z,\bar z) \mapsto A^{\epsilon}_{f;s}&= \int_{\C}\frac{d^2z}{\pi}\, e^{-\vert z\vert^2}\, f(z,\bar z)\,  \mathcal{N}_s(\vert z\vert^2) \, | z;s, {\epsilon} \rg \lg z;s, {\epsilon} |\\
&= \sum_{n,n'} \left \lbrack A^{\epsilon}_{f;s}\right\rbrack_{nn'}\,  |e^{\epsilon}_n;s\rg\lg e^{\epsilon}_n;s|\, ,
\end{align}
where the matrix elements of the operator $ A^{\epsilon}_{f;s}$ are (at least formally) given by the integral:
\begin{align}
\label{matelAf}
\nonumber \left \lbrack A^{\epsilon}_{f;s}\right\rbrack_{nn'} &= \frac{s!}{\sqrt{(n+s)!\,(n'+s)!}}\, \int_0^{\infty}du\,e^{-u} \, L_s^{(n)}(u)\, L_s^{(n')}(u)\, u^{\frac{n+n'}{2}}\,\times \\
&\times  \frac{1}{2\pi}\int_0^{2\pi} d\theta\, e^{\pm i(n-n')\theta}\, F(u,\theta)\, ,
\end{align}
with $F(u,\theta) \equiv f(z,\bar z)$, $z\equiv \sqrt{u}\, e^{i \theta}$ and ``+'' (resp. ``-'') is for ``L'' (resp.``R'').

Practical calculations concern mainly the CS quantization of  elementary blocks of the form $f(z,\bar z) = z^a\, \bar z^b = u^{\frac{a+b}{2}}\, e^{i(a-b)\theta}$. The matrix elements of the corresponding operator are given in Appendix \ref{lagint} together with a few simple cases.
 In particular, for the elementary  functions $f(z, \bar{z}) = z$ and $\bar{z}$ we get  with the help of Eqs. (\ref{eqI5}) and (2.20), (2.19.23.6) in \cite{Prud1}:
\begin{equation}\label{eqCS3}
A^L_{z;s} \,=\, \sum_{n=0}^{\infty}\, \sqrt{s+n+1}\, |e^L_n;s\rangle\langle e^L_{n+1}; s|= A^{L \dag}_{\bar z;s}\, ,
\end{equation}
\begin{equation}\label{eqCS3R}
A^R_{z;s} \,=\, \sum_{n=0}^{\infty}\, \sqrt{s+n+1}\, |e^R_{n+1}; s \rangle \langle e^R_n ;s|= A^{R \dag}_{\bar z;s}\,\,.
\end{equation}
We note that  $A^L_{z;s}$ is lowering operator for the basis $\{ |e^L_n;s\rangle\, , \, n \in \N\}$ of $\mathcal{H}^{L}_{s}$,  with $A^L_{z;s}|e^L_0;s\rangle = 0$, whereas $A^{R \dag}_{\bar z;s}$  is lowering operator for the basis $\{ |e^R_n;s\rangle\, , \, n \in \N\}$ of $\mathcal{H}^{R}_{s}$,  with $A^R_{z;s}|e^R_0;s\rangle = 0$. With the choice $\mathcal{H}^{\epsilon}_{s} = \mathcal{K}^{\epsilon}_{s}$ these operators with their adjoints  are related to  the operators $A^{\epsilon}$ in (\ref{ALRK}) by
\begin{equation}
\label{projALRK}
A^L_{z;s} = \Pi^L_{s}\, A^L \, \Pi^L_{s}\, , \  A^R_{\bar z;s} = \Pi^R_{s}\, A^R\, \Pi^R_{s}\, ,
\end{equation}
where ${\Pi^{\epsilon}}_{s}$ is the orthogonal projector on subspace $\mathcal{K}^{\epsilon}_{s}$.
The lowering $A^L_{z;s}$ (resp. $A^{R \dag}_{\bar z;s}$) and raising $A^L_{\bar{z};s}$ (resp. $A^R_{z;s}$) operators fulfill the commutation relations
\begin{eqnarray}\nonumber
\left[A^L_{z;s}, A^L_{\bar{z};s}\right] &=& \sum_{n=0}^{\infty}\, (n+s+1) \, \left(|e^L_n;s\rangle\langle e^L_n;s| - |e^L_{n+1};s\rangle\langle e^L_{n+1}, s|\right), \\
&=& \mbox{\large 1}_{\mathcal{H}^L_{s}} + s |e^L_0;s\rangle\langle e^L_0;s|\, . \label{eqCS4}
\end{eqnarray}
\begin{eqnarray}\nonumber
\left[ A^R_{\bar{z};s},A^R_{z;s}\right] &=& \sum_{n=0}^{\infty}\, (n+s+1) \, \left(|e^R_n;s\rangle\langle e^R_n;s| - |e^R_{n+1};s\rangle\langle e^R_{n+1}, s|\right), \\
&=& \mbox{\large 1}_{\mathcal{H}^R_{s}} + s |e^R_0;s\rangle\langle e^R_0;s|. \label{eqCS4R}
\end{eqnarray}
The case $s=0$, $\epsilon=L$ yields the  canonical commutation rule for $A^L_{z;s}$, $A^L_{\bar{z};s}$, this is, $[A^L_{z;s}, A^L_{\bar{z};s}] = \mbox{\large 1}_{\mathcal{H}^L_{0}}$. For non-zero values of $s$, there is an extra term proportional to the orthogonal projector on the ``ground state''  $|e^L_0;s\rangle$. The sign of the commutator is reversed under the mirror symmetry $L\mapsto R$.

The position $Q^{\epsilon}_s \equiv A^{\epsilon}_{q;s}$ and momentum $P^{\epsilon}_s \equiv A^{\epsilon}_{p;s}$ operators are  easily obtained by linearity from  the relations $q = (z + \bar{z})/\sqrt{2}$, $p = -i(z - \bar{z})/\sqrt{2}$. With the help of eqs. (\ref{eqCS3}) we obtain:
\begin{equation}\label{eqCS5}
Q^{\epsilon}_s \,=\, \sum_{n=0}^{\infty} \sqrt{\frac{s+n+1}{2}}\, \left(|e^{\epsilon}_n;s\rangle\langle e^{\epsilon}_{n+1};s| + | e^{\epsilon}_{n+1};s\rangle\langle  e^{\epsilon}_{n};s|\right)\, ,
\end{equation}
\begin{equation}\label{eqCS6}
P^{\epsilon}_s \,=\, (-1)^{\epsilon}i\sum_{n=0}^{\infty} \sqrt{\frac{s+n+1}{2}}\, \left(|e^{\epsilon}_{n};s\rangle\langle e^{\epsilon}_{n+1};s| - |e^{\epsilon}_{n+1}; s\rangle\langle e^{\epsilon}_{n};s|\right)\,,
\end{equation}
where $(-1)^{\epsilon} = -1$ for ``L'' and $=1$ for ``R''.
The matrix form of operator $Q^{\epsilon}_s$ is of the symmetric Jacobi type:
\begin{equation}
\label{QJacobi}
Q^{\epsilon}_s = \left(
\begin{array}{llllll}
    0 & \sqrt{\frac{s+1}{2}} & 0 & \cdots \\
    \sqrt{\frac{s+1}{2}} & 0 & \sqrt{\frac{s+2}{2}} & \cdots \\
    0 & \sqrt{\frac{s+2}{2}} & 0 & \ddots  \\
    \vdots &  & \ddots & \ddots \\
\end{array}\right)\, ,
\end{equation}
We have a similar expression for $P^{\epsilon}_s$. Their commutator,
\begin{equation}
\label{hermscr}
[Q^{\epsilon}_s, P^{\epsilon}_s] = i[A^{\epsilon}_{z;s}, A^{\epsilon}_{\bar{z;s}}] = (-1)^{\epsilon+1}i(\mbox{\large 1}_{\mathcal{H}^L_{s}} + s |e^{\epsilon}_0;s\rangle\langle e^{\epsilon}_0;s|)
\end{equation}
 is ``almost''canonical, in the sense that  like for (\ref{eqCS4}) there is the  extra projector on the ground state multiplied by  $i\,s$.


\section{Quantum localization and  associated orthogonal polynomials}
\label{chihara}
In this section, we analyze in more details, like in \cite{alietal09},  the localization properties of the ``almost'' canonical operators (\ref{eqCS5}) and (\ref{eqCS6}).
Actually, it is enough to study the quantum position operator $Q^{\epsilon}_s $.  Henceforth, for a sake of simplicity, we denote it by $Q$ and we denote the basis element $| e^{\epsilon}_{n};s\rangle$ by $|e_n\rg$. We also put $x_n = s+n$.
The operator $Q$ acts on the basis vectors $|e_n\rg  $ in the manner,
\begin{equation}
  Q|e_{n}\rg = \sqrt{\frac{x_n}2}\; |e_{n-1}\rg+
  \sqrt{\frac{x_{n+1}}2}\; |e_{n+1}\rg\; .
\label{eq:pos-op-act}
\end{equation}
If the  sequence $(x_n)$ is such that the sum $\displaystyle{\sum_{n=0}^\infty
\frac 1{\sqrt{x_n}}}$ diverge, then the operator $Q$ is essentially
self-adjoint \cite{akhiezer65,chihara78,simon96} and hence has a
unique self-adjoint extension, which we again
denote by $Q$. It is precisely the case with $x_n= s+n$. Let $E_\lambda , \;
 \lambda \in \mathbb R$, be the spectral family of $Q$, so that,
$$ Q = \int_{-\infty}^\infty \lambda \; dE_\lambda \; .$$
Thus there is a measure $ \varpi(d\lambda)$ on $\mathbb R$ such that on the
Hilbert space $L^2 (\mathbb R , \varpi)$,
the action of
$Q$ is just a multiplication by $\lambda$ and the basis vectors
$|e_{n}\rg$ can  be represented by functions
$p_n (\lambda )$ (see \cite{askeyismail84} for instance and references there). Consequently, on this space, the relation (\ref{eq:pos-op-act})
assumes the form
\begin{equation}
  \lambda p_n (\lambda) = c_n  p_{n-1}(\lambda ) + c_{n+1} p_{n+1} (\lambda)\; ,
   \qquad c_n =  \sqrt{\frac{x_n}2}\;,
\label{eq:pos-op-act2}
\end{equation}
which is a two-term recursion relation, familiar from the theory of orthogonal
polynomials. It follows that
$\varpi (d\lambda ) = d(\lg e_0|E_\lambda|e_0\rg $, and the $p_n$
may be realized as the polynomials obtained
by orthonormalizing the sequence of monomials $1, \lambda, \lambda^2 , \lambda^3 ,
\ldots\; , $ with respect to this measure
(using a Gram-Schmidt procedure).  Furthermore, for any $\varpi$-measurable set
$\Delta \subset \mathbb R$,
\begin{equation}
\lg e_n| E(\Delta)| e_m\rg = \int_{\Delta}  p_n (\lambda )p_m (\lambda )\;\varpi (d\lambda )\; ,
\label{eq:poly-basis}
\end{equation}
and
\begin{equation}
   (p_n, p_m)_{L^2 (\mathbb R , \varpi )} = \int_{\mathbb R}
       p_n (\lambda )p_m (\lambda )\; \varpi (d\lambda ) = \delta_{m n}\; .
\label{eq:poly-orthog}
\end{equation}

The polynomials $p_n$ are not {\em monic polynomials\/}, i.e., that the coefficient of $\lambda^n$ in
$p_n$ is not one. However, the renormalized polynomials
\begin{equation}
  q_n (\lambda ) = {c_n !}\; p_n (\lambda ) \, ,  \  c_n! \deq c_1c_2\cdots c_n\, , \ c_0!=1\, ,
\end{equation}
are seen to satisfy the recursion relation
\begin{equation}
  q_{n + 1} (\lambda ) = \lambda\; q_n (\lambda )  - c_n^2\; q_{n-1} (\lambda )\; ,
\label{eq:mon-rec-reln}
\end{equation}
from which it is clear that these polynomials are indeed monic.

   There is a simple way to compute the monic polynomials.
Let $Q_n$ be the truncated matrix consisting of the first $n$ rows and columns of $Q$ in (\ref{QJacobi}) and $\lu_n$
the $n\times n$ identity matrix. Then,
\begin{equation}
  \lambda \lu_n - Q_n =
\begin{pmatrix} \lambda &  - c_1 & 0 & 0 & 0 &\ldots & 0 & 0 & 0\\
- c_1 & \lambda & - c_2 & 0 & 0 & \ldots & 0 & 0 & 0\\
0 & - c_2 & \lambda & - c_3 & 0 & \ldots & 0 & 0 & 0\\
0 & 0 & - c_3 & \lambda & - c_4 & \ldots & 0 & 0 & 0 \\
0 & 0 & 0 & - c_4 & \lambda & \ldots & 0 &0 & 0\\
\vdots & \vdots &\vdots &\vdots &\vdots &\ddots & \vdots & \vdots & \vdots\\
0 & 0 & 0 & 0 & 0 & \ldots & \lambda & -c_{n-2} & 0\\
0 & 0 & 0 & 0 & 0 & \ldots & -c_{n-2} & \lambda & -c_{n-1}\\
0 & 0 & 0 & 0 & 0 & \ldots & 0 & - c_{n-1} & \lambda \end{pmatrix}\; .
\end{equation}
It now follows that $q_n$ is just the characteristic polynomial of $Q_n$ :
\begin{equation}
  q_n (\lambda ) = \text{det} [ \lambda \lu_n - Q_n ]\; .
\end{equation}
Indeed, expanding the determinant with respect to the last row, starting at the lower right
corner, we easily get
\begin{equation}
\text{det} [ \lambda \lu_n - Q_n ] = \lambda \;\text{det} [ \lambda \lu_{n-1} - Q_{n-1} ] -
    c_{n-1}^2 \;\text{det} [ \lambda \lu_{n-2} - Q_{n-2} ]\; ,
\end{equation}
which is precisely the recursion relation (\ref{eq:mon-rec-reln}). Consequently the roots of the polynomial
$q_n$ (or $p_n$) are the eigenvalues of $Q_n$.

In the present context where $c_n = \sqrt{x_n/2}= \sqrt{(n+s)/2}$, it is  convenient to make explicit the parameter $s$ and to work with non-monic polynomials $H_n(x;s)$
deduced from the $p_n$'s as
\begin{equation}
H_n(\lambda;s) = 2^{n} \,q_n (\lambda) = 2^{n} c_n!\,p_n (\lambda)\, ,
\label{redpol}
\end{equation}
so that the recurrence relation (\ref{eq:pos-op-act2}) now reads
\begin{align}
H_{n+1} (\lambda;s) &= 2 \lambda H_n (\lambda;s) - 4c_n^2   H_{n-1}(\lambda; s )= 2 \lambda H_n (\lambda;s) - 2(s+n)   \tilde H_{n-1}(\lambda;s )\,,
 \\  \nonumber \tilde H_{-1;s} &= 0\, , \quad \tilde H_0(\lambda;s) = \lu\, .
\label{eq:pos-op-act2}
\end{align}
Indeed, in the case $s= 0$, with such initial conditions, this recurrence is solved by Hermite polynomials, $H_n(\lambda;0) = H_n(\lambda)$ \cite{magnus66}. For a general value of $s$, not necessarily integer, these polynomials are named \textit{associated Hermite polynomials}. They were introduced and studied by Askey and Wimp in \cite{askeywimp84} and were completely characterized by Ismail, Letessier and Valent in \cite{ismailetal88}.
Askey and Wimp \cite{askeywimp84} solved  the spectral measure problem through the explicit orthogonality relation:
\begin{equation}
\label{asshermorth}
\int_{-\infty}^{\infty} \frac{H_m (\lambda;s)\,H_n (\lambda;s)}{\vert D_{-s}(\lambda e^{i\pi/2}\sqrt 2\vert^2}\,d\lambda = 2^n\sqrt{\pi}\Gamma(n+s+1)\delta_{mn}\, ,
\end{equation}
where $D_{\nu}$ is a parabolic cylinder function\cite{magnus66}. Their expression was given in \cite{ismailetal88} in terms of associated Laguerre polynomials,
\begin{align}
\label{asshermlagev}
 H_{2n}(\lambda; s )&= \sigma_n \mathcal{L}_n^{-1/2}(\lambda^2;s/2)\, ,
 \\ \label{asshermlagodd} H_{2n+1}(\lambda; s )&= 2\lambda \sigma_n L_n^{1/2}(\lambda^2;s/2)\, ,\quad \sigma_n= (-4)^n(1+s/2)_n\,.
\end{align}
These polynomials  are given by:
\begin{align}
\label{lagassI} \mathcal{L}_n^{\alpha}(x;c) &= \frac{(\alpha+1)_n}{n!}\sum_{m=0}^n \frac{(-n)_mx^m}{(c+1)_m(\alpha+1)_m}\, {}_3F_2(m-n,m-\alpha,c;-\alpha -n,c+m+1;1)
  \, ,  \\
\label{lagassII} L_n^{\alpha}(x;c)&= \frac{(\alpha+1)_n}{n!}\sum_{m=0}^n \frac{(-n)_mx^m}{(c+1)_m(\alpha+1)_m}\, {}_3F_2(m-n,m+1-\alpha,c;-\alpha -n,c+m+1;1)\,.
\end{align}

\section{The question of equivalence  of quantizations  in regard to the harmonic oscillator }
\label{canvcs}

We have seen  that the $s$-dependent CS quantization of the classical position $q$ provides a position operator $Q$: $A_q = Q$ and a momentum operator $P=A_p$. They are both essentially self-adjoint and the former acts as the multiplication operator $ Q \psi(x) = x \psi(x)$ when it is realized on the Hilbert space of ``spatial'' wave functions $\psi \in L_{\C}^2(\R, \varpi(dx))$ defined as square integrable functions on its own spectrum $\R$.
Due to the non-canonical commutation rule (\ref{hermscr}) we cannot expect that for $s\neq 0$ $P$ acts as the simple derivation $\mp id/dx$ on this space. Its operatorial expression could be quite involved (see e.g. \cite{borzov01}).
  Let us now compare the  operator $Q^2 = \left(A_q\right)^2$ (resp. $P^2 = \left(A_p\right)^2$) with $A_{q^2}$ (resp. $A_{p^2}$), the CS quantized of the square of the classical position (reps. momentum). A simple calculation  based on the direct  squaring of (\ref{eqCS5}) (reps.(\ref{eqCS6})) on one hand and on the  computation of $A_{q^2}$ (resp. $A_{p^2}$) based on the relation $q^2 = \vert z\vert^2 + (z^2 + \bar z^2)/2)$ (resp. $p^2 = \vert z\vert^2 - (z^2 + \bar z^2)/2$) and the integral (\ref{monquant}) given in Appendix \ref{lagint} on the other hand  yields:
\begin{align}
\label{Q2Aq2}
\nonumber A_{q^2} &= \sum_{n=0}^{\infty}(n+2s +1) |e_n\rg \lg e_n|   + \sum_{n=0}^{\infty}c_{n+1}c_{n+2} \left(|e_n\rg\lg e_{n+2}| + |e_{n+2}\rg\lg e_{n}| \right) \\ &= Q^2 + \left( s+ \frac{1}{2}\right)\lu + \frac{s}{2}|e_0\rg\lg e_0|\, .
\end{align}
\begin{align}
\label{P2Ap2}
\nonumber A_{p^2} &= \sum_{n=0}^{\infty}(n+2s +1) |e_n\rg \lg e_n|   - \sum_{n=0}^{\infty}c_{n+1}c_{n+2} \left(|e_n\rg\lg e_{n+2}| + |e_{n+2}\rg\lg e_{n}| \right) \\ &= P^2 + \left( s+ \frac{1}{2}\right)\lu + \frac{s}{2}|e_0\rg\lg e_0|\, .
 \end{align}
  Therefore,  the  operators $Q^2$ and $A_{q^2}$ (resp. $P^2$ and $A_{p^2}$) differ by a multiple  the unity plus a multiple of the orthogonal projector on the lowest state, and, consequently, their respective spectra differ. The spectrum of $Q^2$ (resp. $P^2$) is $\R^{+}$ with infimum $0$.  Therefore, the infimum of the spectrum of $A_{q^2}$ (resp. $A_{p^2}$) exists since it is positive and is larger or equal to $s+1/2$ from the inequality:
  \begin{equation}
\label{infimumAq2}
\underset{\psi\in D(Q^2), \|\psi\|=1}{\inf}\lg \psi|A_{q^2}|\psi\rg\geq\underset{\psi\in D(Q^2), \|\psi\|=1}{\inf}\lg \psi|Q^2|\psi\rg+\frac{s}{2}\vert\lg e_0|\psi \rg\vert^2
+s+\frac{1}{2}\, .
\end{equation}

Now, let us choose for $\psi$ a coherent state $|z/\sqrt\sigma;s;\epsilon\rg \equiv |z/\sqrt\sigma\rg$, given by Eqs(\ref{eqCS0}-\ref{eqCS1R}), where we have rescaled the variable $z$ with a square-rooted width parameter.  As $\sigma \to 0$, we expect that the lower symbol $\lg z/\sqrt\sigma|Q^2| z/\sqrt\sigma\rg$ and the probability $\vert\lg e_0|z/\sqrt\sigma \rg\vert^2= \left(L_s^{(0)}(\vert z\vert^{2}/\sigma)\right)^2/\mathcal{N}_{s}(\vert z\vert^{2}/\sigma)$ concentrates to a  peak localized at the origin, due to the dominant exponential term in  $\mathcal{N}_{s}$ as given by Eq.(\ref{behN(t)}). It follows that
  \begin{equation}
\label{infimumAq2exact}
\underset{\psi\in D(Q^2), \|\psi\|=1}{\inf}\lg \psi|A_{q^2}|\psi\rg =
s+\frac{1}{2}\, ,
\end{equation}
and the same holds for $\underset{\psi\in D(P^2), \|\psi\|=1}{\inf}\lg \psi|A_{p^2}|\psi\rg$.

We now turn our attention to the energy operator for the one-dimensional quantum harmonic oscillator. There are at least  two expressions for it according to the followed quantization scheme.  The first one which appears to us as the most natural is issued from the CS quantization of   the classical Hamiltonian,
$H \,=\, (q^{2} + p^{2})/2 = \vert z \vert^2$. Its quantum version  $A^{\epsilon}_{\vert z\vert^{2}; s}$ is easily calculated and reads as the diagonal operator
\begin{equation}\label{eqCS7}
A_{H} = \frac{1}{2}(A_{p^{2}}  + A_{q^2})=  \sum_{n=0}^{\infty}\, (n+ 2s + 1)\, |e_n \rangle\langle e_n|\, .
\end{equation}
This entails  that the lowest state $|e_0\rangle$ has energy  $(2s + 1)$ and  that the energy levels are equidistant by 1, like for the energy levels of the standard (canonical) case.

The alternative to this direct CS quantization is to use the usual ansatz (as is done in  \cite{borzov01} and related references) which consists in replacing $q$ by $Q$ and $p$ by $P$ in the expression of the classical observable $H \,=\, (q^{2} + p^{2})/2$. This leads to the operator
\begin{equation}\label{eqCS8}
\hat{H} = (P^{2} + Q^{2})/2 = \frac{s\!+\!1}{2}\, |e_0\rangle\langle e_0| + \sum_{n\geq 1}\, (n \!+\! s\!+\! 1/2)\, |e_n\rangle\langle e_n|.
\end{equation}
The distance between the  first and second level is $s/2 + 1$, whereas the distance between the upper levels (e.g., third and  second level and so on) is constant and equal to 1. It is obvious that for $s=0$ Eqs. (\ref{eqCS7}) and (\ref{eqCS8}) differ just by a global shift of $1/2$. The distinctions between them hold for  $s\geq 1$, for which there is  a shift of the ground state energy. Let us consider as granted that in the present case the quantum kinetic energy is $P^2/2$, which is actually the case for $s=0$. Therefore the term $Q^{2}/2$ in (\ref{eqCS8}) is the quantum potential energy if we follow this usual quantization procedure. Now, from the relation between the two quantum Hamiltonians obtained from (\ref{Q2Aq2}) and (\ref{P2Ap2})
\begin{equation}
\label{Hcancs}
A_H= \hat{H} + \left( s+ \frac{1}{2}\right)\lu + \frac{s}{2}|e_0\rg\lg e_0| = \frac{P^{2}}{2} +  \frac{Q^{2}}{2} + \left( s+ \frac{1}{2}\right)\lu + \frac{s}{2}|e_0\rg\lg e_0|\, ,
\end{equation}
we observe that the quantum harmonic potential energy yielded by CS quantization differs from $\hat{H}$ by the diagonal operator $\left( s+ \frac{1}{2}\right)\lu + \frac{s}{2}|e_0\rg\lg e_0|$.  Clearly, for any $s$, CS quantization and canonical-like quantization are not mathematically equivalent. Let us now examine if this non-equivalence is alleviated if we examine difference from a physically more-oriented point of view.

Usually in Physics the zero-point energy is taken at the minimum of the potential energy. In the present case, the quantum potential energy is $  Q^{2}/2 + \left( s+ \frac{1}{2}\right)\lu + \frac{s}{2}|e_0\rg\lg e_0|$ and, following the discussion yielding (\ref{infimumAq2exact}), its infimum is $s+1/2$.  Therefore,
the difference between the ground state  energy of $A_H$ and the zero-point energy is $s+1/2$  and it is not, \textbf{at the exception of the standard case $s=0$}, the same as $(s+1)/2$,  which is the difference between the ground state  energy of $\hat H$ and the value 0 corresponding to the zero-point energy in this case.

So far we did not take into account physical parameters.   In order to define harmonic coherent states $|\xi_{q,p}\rg $ that live on the physical classical phase space $\mathcal{P}=\{ (q,p) \in \mathbb{R}^2 \}$ we need to introduce an arbitrary length scale $\ell$ and the reduced Planck constant $\hbar$. Then we define the normalized vectors $|\xi_{q,p}\rg $ from the states $|z\rg$ in  (\ref{eqCS1}) as
\begin{equation}
|\xi_{q,p}\rg\equiv  \left|z=\frac{q}{\ell \sqrt{2} } + i \frac{p \ell}{ \hbar \sqrt{2}} \right\rg\, .
\end{equation}
The resolution of unity in (\ref{resunity1}) becomes
\begin{equation}
\int_{\mathcal{P}} \frac{dq dp}{2 \pi \hbar} |\xi_{q,p}\rg\lg \xi_{q,p}| = \lu\, .
\end{equation}

   The CS quantization of the classical observables $q$ and $p$ leads to $Q= A_q=\frac{\ell}{\sqrt{2}}(A_z +A_{\bar z})$ and $P=A_p=\frac{\hbar}{i \ell \sqrt{2}} (A_z- A_{\bar z})$. Operators $P$ and $Q$ verify the commutation rule (we restrict the discussion  to the ``$L$'' case) $[Q,P]=i \hbar \lu + i\hbar |e_0\rg\lg e_0|$.

   At this stage $\ell$ is a free parameter of the theory, since, on a physical point of view, only the spectra of  the operators $P$ and $Q$ are observables.
 We now introduce the mass $m$ of the particle (or the reduced mass of two particles). The quantized kinetic energy $A_{p^2/2m}$ is
\begin{equation}
A_{p^2/2m}=\frac{P^2}{2m}+ \frac{\hbar^2}{4 m \ell^2} ((2s + 1)\lu +  s |e_0\rg\lg e_0|)\, .
\end{equation}
The additive term  must be  viewed as an ``internal energy" operator similar to the $mc^2$ term appearing in a relativistic approach. In fact, if we decide to fix $\ell$ as being one-half of the Compton length $\ell=\frac{\hbar}{2m c}$ associated with the mass $m$, we obtain exactly $\frac{\hbar^2}{4 m \ell^2}=m c^2$ as the factor giving the physical dimension to this term.

   Furthermore the classical harmonic potential is $V(q)= \frac{1}{2} k q^2 \equiv \frac{1}{2} m \omega^2 q^2$ where $k$ is the constant force and $\omega$ is the usual vibrational parameter.  The CS quantized counterpart $A_{V(q)}$ of $V(q)$ reads as
\begin{equation}
A_{V(q)}= \frac{1}{2} m \omega^2 Q^2 +\frac{1}{4} m \omega^2 \ell^2 ((2s + 1)\lu +  s |e_0\rg\lg e_0|)\,.
\end{equation}
   Hence, the CS quantization of the classical hamiltonian $H(p,q)=p^2/2m+V(q)$ leads to the quantum hamiltonian
\begin{equation}
A_H= \frac{P^2}{2m} + \frac{1}{2} m \omega^2 Q^2 + \left(\frac{\hbar^2}{4 m \ell^2}+\frac{1}{4} m \omega^2 \ell^2\right) ((2s + 1)\lu +  s |e_0\rg\lg e_0|)\, .
\end{equation}
   If we choose (as it was previously done) the free parameter $\ell$ as the one-half of the Compton length $\ell=\frac{\hbar}{2m c}$, we obtain
\begin{equation}
A_H=  \frac{P^2}{2m} + \frac{1}{2} m \omega^2 Q^2 + (mc^2+ \gamma\hbar \omega)((2s + 1)\lu +  s |e_0\rg\lg e_0|)\, .
\end{equation}
    Here  $\gamma = \frac{\hbar \omega}{16 mc^2}$ is a dimensionless factor expressing the ratio between two typical energies of the model, namely the (non-relativistic) quantum energy $\hbar \omega$ and the rest mass of the particle.

   Since the validity of the classical hamiltonian $H(p,q)$ is restricted to the non-relativistic domain, and since in this case the ratio $\gamma$ is completely negligible, we obtain
\begin{equation}
A_H\simeq \frac{P^2}{2m} + \frac{1}{2} m \omega^2 Q^2 + mc^2((2s + 1)\lu +  s |e_0\rg\lg e_0|)\,.
\end{equation}
    This is  the quantum hamiltonian yielded by the usual quantization ansatz, up to the very large additional term proportional to $mc^2$, i.e. a sort of quantum  energy operator proper to and only to  the particle. Hence, if  we can assert that in the standard case $s=0$ CS and canonical quantizations are \emph{physically} equivalent as far as we are concerned with harmonic vibrations, it is not possible to pretend that such an equivalence holds  in the other cases.

%


\section{Conclusion}
\label{seconc}
In this paper we have explored an infinite family of possible quantizations of the complex plane based on the existence of complex Hermite coherent states. The complex plane can be viewed as the phase space for the motion of a particle on the line. It could   be as well viewed as the plane of quadratures in electromagnetism. It could represent something more unusual or exotic. Whatever the  interpretation one can have of it  in mathematics (e.g. reproducing kernel spaces), in physics (e.g. non standard quantizations of classical vibrations, non commutative quantum mechanics) or in signal processing,  our analysis is aimed to cast more interest in the existence  of these various ``quantum fashions'' of analyzing a ``classical object''. This leads to the intriguing question of equivalence or not between such different approaches.

In the continuation of our work, which is restricted here to \textit{irreducible} quantizations,  an interesting idea would be  to explore \textit{reducible} ones by dealing with quantizations based on the finite sum of subspaces $\mathcal{K}^{\epsilon}_{s}$ in $L^2( \C, d^2z/\pi)$,
\begin{equation}
\label{finorthdec}
\mathcal{L}_S \deq \bigoplus_{s=0}^{S-1} \left(  \mathcal{K}^L_{\ast s}\oplus  G_s\oplus \mathcal{K}^R_{\ast s}\right) \, ,
\end{equation}
and to examine issues in its  mathematical and physical aspects, for instance by analyzing the behavior of lower symbols of the commutator $[A_q,A_p]$.

Other avenues to explore could be
supersymmetric quantum mechanics and modular von Neumann algebraic structures in the present context.

\section*{Acknowledgments}
The authors are very indebted to Mourad E.H. Ismail for valuable informations on associated Laguerre and Hermite polynomials. They would also like to acknowledge financial support from the
   Universit\`a di Palermo through Bando CORI, cap. B.U. 9.3.0002.0001.0001.
One of us (STA) would like to acknowledge a grant from the Natural Sciences and Engineering Research Council (NSERC) of Canada.
\appendix
\section{Generalized Laguerre polynomials}
\begin{equation}
\label{laguerre}
L_s^{(\alpha)}(x) = \sum_{m=0}^s (-1)^m \, \binom{n+\alpha}{n-m}\, \frac{x^m}{m!}\, .
\end{equation}

%

\section{Laguerre integrals and matrix elements}
\label{lagint}
\paragraph{Orthogonalty}
\begin{equation}
\label{ortholag}
\int_0^{\infty}dx\, e^{-x}\, x^{\alpha}\, L_m^{(\alpha)}(x)\,  L_n^{(\alpha)}(x)= \delta_{mn}\, \Gamma(1+\alpha) \, \binom{n+\alpha}{n}\, .
\end{equation}
\paragraph{One Laguerre polynomial and simple power}
\begin{equation}
\label{ortholag2}
\int_0^{\infty}dx\, e^{-x}\, x^{\lambda}\, L_n^{(\alpha)}(x) = \frac{\Gamma(\lambda + 1) \, \Gamma(\alpha-\lambda + n)}{n!\, \Gamma(\alpha-\lambda)}\, .
\end{equation}
\paragraph{
Two Laguerre polynomials and simple power \cite{Prud1}:}
\begin{align}
\label{lagintegr}
\nonumber &\int_0^{\infty} dx\, x^{\lambda}\, e^{-x}\, L_r^{(\alpha)}(x)\, L_{s}^{(\beta)}(x)
 \\ \nonumber& =\frac{(1+\alpha)_r\,(\beta -\lambda)_s\,\Gamma(\lambda +1)}{r!\, s!}\, {}_3F_2(-r,\lambda +1,\lambda +1 -\beta;\alpha +1,\lambda +1-\beta   -s;1)\\ & = \frac{(1+\beta)_r\,(\alpha -\lambda)_s\,\Gamma(\lambda +1)}{r!\, s!}\, {}_3F_2(-s,\lambda +1,\lambda +1 -\alpha;\beta +1,\lambda +1-\alpha  -r;1)\, , \\ \nonumber &\hskip 9cm \Re\lambda > -1\, .
\end{align}
From this integral is easily derived the expression of the matrix elements of the CS quantized version of the monomial $z^a\,\bar z^b$, $a,\, b \in \N$:

\begin{align}
\label{monquant}
\nonumber &\left \lbrack A^{\epsilon}_{z^a\,\bar z^b;s}\right\rbrack_{nn'} = \frac{s!}{\sqrt{(n+s)!\,(n'+s)!}}\, \int_0^{\infty}du\,  e^{-u}\, L_s^{(n)}(u)\, L_s^{(n')}(u)\, u^{\frac{n+ a+ n' +b}{2}}\times \\
\nonumber
 &\hskip 6cm \times \, \frac{1}{2\pi} \int_0^{2\pi}  d\theta\, e^{ i[\epsilon(n-n') + a-b]\theta}\\
\nonumber &= \delta_{n- n', \epsilon(b -a)}\frac{s!}{\sqrt{(n+s)!\,(n'+s)!}}\, \int_0^{\infty}du\,  u^{n+a_{\epsilon}}\,  e^{-u}\, L_s^{(n)}(u)\, L_s^{(n')}(u)\\
&= (-1)^s \delta_{n- n', \epsilon(b -a)}\,  \sqrt{\frac{(n+s)!}{(n'+s)!}}\,\sum_{m=\sup(0,s-a_{-\epsilon})}^s(-1)^m\, \frac{(n+a_{\epsilon}+m)!\, ( a_{-\epsilon}+m)!}{m!\, (s-m)!\,(m+n)!\, (a_{-\epsilon} -s+m)!}\\
\nonumber &= (-1)^s \delta_{n- n', \epsilon(b -a)}\,  \sqrt{\frac{(n+s)!}{(n'+s)!}}\,\frac{a_{-\epsilon}!\, (n+ a_{\epsilon})!}{n!\, s!\, (a_{-\epsilon} -s)!}\times \\
\label{monquant1}& \times {}_3F_2(-s, n + a_{\epsilon} +1 , a_{-\epsilon} +1; n +1,  a_{-\epsilon}  - s +1;1)\, .
\end{align}
Here, $\epsilon$ stands for $L \equiv +$ or $R \equiv -$, and $a_+ \equiv a$, $a_- \equiv b$, .

%

\end{document}